\documentclass{aa} 
\usepackage{aalongtable}
\usepackage{aalongtable,lscape}
\usepackage[figuresright]{rotating}
\usepackage{epsfig}
\usepackage{graphicx}
\usepackage{natbib}
\bibpunct{(}{)}{;}{a}{}{,} 
\usepackage{txfonts}
\def\sun{\hbox{$\odot$}}
\def\deg{^\textrm{\scriptsize o}}
\begin{document}
\title{Water maser detections in southern candidates to post-AGB stars and Planetary Nebulae}

\author{O. Su\'arez \inst{1,2}
        \and J.F. G\'omez \inst{2}
        \and L.F. Miranda \inst{2}
        \and J.M. Torrelles \inst{3}
        \and Y. G\'omez \inst{4}
        \and G. Anglada \inst{2}
        \and O. Morata \inst{5,6,7}
                }

\offprints{O. Su\'arez}

\institute{UMR 6525 H.Fizeau, Universit\'e de Nice Sophia Antipolis, CNRS, OCA. Parc
  Valrose, F-06108 Nice Cedex 2, France. olga.suarez@unice.fr
\and
Instituto de Astrof\'{\i}sica de Andaluc\'{\i}a, CSIC, Apartado
  3004, E-18080 Granada, Spain
\and
Instituto de Ciencias de Espacio (CSIC)-IEEC, Facultat de F\'{\i}sica,
Planta 7a, Universitat de Barcelona, Mart\'{\i} i Franqu\`es  1, 08028
Barcelona, Spain
\and
Centro de Radioastronom\'{\i}a y Astrof\'{\i}sica, UNAM, Campus
Morelia, Apdo. Postal 3-72, Morelia, Michoac\'an 58089, Mexico
\and
Department of Earth Sciences, National Taiwan Normal University, 88 Sec.4, Ting Chou Rd., Taipei 116, Taiwan
\and
Academia Sinica Institute of Astronomy and Astrophysics, P.O. Box 23-141, Taipei 10617, Taiwan
\and
Departament d'Astronom\'{\i}a i Meteorolog\'{\i}a, Universitat de
Barcelona, Mart\'{\i} i Franqu\`es  1, 08028 Barcelona, Spain
}
\date{Received <date>/Accepted <date>}

\abstract
   {} 
{We intended to study the incidence and characteristics of water masers in the envelopes of stars in
  the post-AGB and PN evolutionary stages.}
{We have used the 64-m antenna in Parkes (Australia)
to search for water maser emission at 22 GHz, towards
      a sample of 74 sources with IRAS colours characteristic of
      post-AGB stars and PNe, at declination $< -32 \deg$. In our sample, 39\% of the sources are PNe or PNe candidates, and 50\% are post-AGB stars or post-AGB candidates.} 
    {We have detected four new water masers, all of them in optically
      obscured sources: three in PNe candidates (IRAS 12405$-$6219, IRAS
      15103$-$5754, and IRAS 16333$-$4807) and one in a post-AGB candidate (IRAS 13500$-$6106). The PN candidate IRAS 15103$-$5754 has water fountain
      characteristics, and it could be the first PN of this class found.}
    {We confirm the tendency suggested in Paper I that the presence of
      water masers  during the post-AGB phase is favoured
      in obscured 
      sources with massive envelopes, in comparison with objects with
      optical counterparts in these phases.
      We propose an evolutionary scenario for water masers in the
      post-AGB and PNe stages, in which ``water fountain'' masers
      could develop during post-AGB and early PN stages. Later PNe
      would show lower velocity maser emission, both along jets and
      close to the central objects, with only the central masers
      remaining in more evolved PNe. }

\keywords{Masers -- Surveys -- Stars: AGB and post-AGB -- Stars: Mass-loss -- Planetary Nebulae}

\maketitle
%
%
\section{Introduction}


The study of H$_2$O masers in evolved stars has been proved to be a
very powerful tool to understand the evolution and shaping of sources 
between the Asymptotic Giant Branch (AGB) and the phase of
Planetary Nebula (PN). 
About 80\% of the stars (those with M$\le$8M$\sun$) go through these
evolutionary stages, where a dramatic change in their morphology is
produced. The stars in the AGB stage have a high mass-loss rate, that
can reach $\simeq 10^{-4}$ M$_{\sun}$yr$^{-1}$. This mass-loss is
usually spherical, although the first manifestations of asymmetry are
already observed in some of them \citep[see for
example][]{leao06,josselin00}. A large fraction of PNe show, however, axial
symmetries with bipolar or multipolar morphologies and, in many cases,
jets \citep{sahai07}.

Water maser emission is frequently present in the AGB
stage, with the spectra showing double-peak emission in some cases
(e.g., \citealt{engels86}),
similar to those of OH masers in this type of sources, which are
characteristic of an
spherical envelope expanding at velocities $\sim$10$-$30~km~s$^{-1}$
\citep{reid77, bowers89}. In AGB stars, we can also find SiO maser
emission, and the three species are distributed in such a way 
that SiO masers are
located close to the central star ($\sim$10 AU), H$_2$O are located at
the inner part of the envelope (between 10 and 100 AU) and OH masers
at distances $> 10^3$ AU from the central star
\citep{reidmoran81}.

In the classical view of maser emission pumped by spherical AGB winds,
SiO masers are expected to survive only $\simeq 10$ yr after the stop
of the strong mass-loss, while H$_2$O and OH masers will be
extinguished after $\simeq 100$ yr and $\simeq 1000$ yr, respectively
\citep{lewis89,gomezy90}. These timescales imply that in the post-AGB
stage we can still find some objects that harbour water masers with
the same characteristics of those found in the AGB stage \citep[see
for example][]{engels02}. On the other hand, water maser emission was
not expected in PNe.

There are cases, however, of evolved sources in which the observed 
water maser
emission departs from this pattern expected from spherical
mass-loss. In several post-AGB stars, called ``water
fountains'', the H$_2$O maser emission traces highly collimated
bipolar jets with a velocity spread $\ga 100$ km~s$^{-1}$, and
dynamical ages $\la 100$ yr \citep{imai07a}, in what could represent the first
manifestation of collimated mass-loss in evolved stars. Moreover,
now we know of three confirmed cases of PNe harbouring water masers
\citep{miranda01, degregorio04, gomez08}. These three ``H$_2$O-PNe'' have in
common a bipolar morphology, suggesting that the asymmetrical mass
loss can be related to the presence of water masers in PNe.

In order to study the possible relationships between the evolutionary
stage of evolved objects and the presence of water masers,
\citet{suarez07} (hereafter Paper I) carried out a single-dish survey
for H$_2$O maser emission towards the evolved sources in the catalogue
of \citet{suarez06} (hereafter SGM) that were located north of
declination $-32\deg$. A total of 105 sources were searched with the
Robledo 70m antenna, with five detections, four of them reported in
that paper for the first time.  Three of these detections were found
in obscured sources, of which one was a water fountain (confirmed by
\citealt{suarez08}) and another one a PN candidate. These results led
to suggest that the presence of water masers in the post-AGB and
  PN phases is favoured in stars with massive envelopes, and that PNe
harbouring water masers are a special class of massive, rapidly
evolving PNe. Water masers in these sources would be related to
processes of non-spherical mass-loss, rather than being the remnant of
water masers pumped by spherical winds in the AGB stage. In Paper I it
was also suggested that probably all H$_2$O-PNe had traversed
previously by a water fountain phase \citep[see also][]{gomez08}. The
discovery of objects showing both water fountain and PNe
characteristics would be an important evidence of this evolutionary
path.

In this paper we present the continuation of the work started in Paper I, with
single-dish observations of sources in the SGM catalogue with declination
$\le-$32$\deg$, using the Parkes antenna.
This paper is structured as follows: 
In Sec. \ref{selection} we present the sample of observed sources. 
In Sec. \ref{observations} we describe the observations performed.
In Sec. \ref{results} we describe the results obtained, including
a description of the detected objects. In Sec. \ref{discussion} we discuss
our results and finally, in Sec. \ref{conclusions} we give
the main conclusions of this work.

\section{Description of the source sample}
\label{selection}

The selection criteria are described in detail in Paper I and in SGM. 
Briefly, the SGM atlas is composed of IRAS
sources located in the region of the IRAS colour-colour diagram
populated mainly by post-AGB stars \citep{vanderveen89}. The
sources in this region would have expanding envelopes characterised by a range of dust
temperatures between 200\,K and 80\,K, and
radii between 0.01\,pc and 0.1\,pc. 

The evolved sources in the SGM catalogue are classified in five
different groups: PN, post-AGB stars, transition sources (that are
evolving between the post-AGB and PN phases), peculiar sources (evolved
sources with peculiar characteristics), and objects with no optical
counterpart (which could be evolved objects in any of the stages mentioned
above, but could no be classified with the optical spectroscopy presented in
SGM). 

For these objects with no optical counterpart found by SGM, we
  can only give a tentative classification as ``post-AGB candidate''
  or ``PN candidate''. We consider them 
candidates to PN when there is an indication of the ionisation of
the envelope, such as the presence of radio continum
emission. In many cases, these PN candidates are merely point-like 
sources in infrared images, with no further evidence of the presence
of a nebula. 
When there is no indication 
of ionisation or nebulosity, we consider these obscured objects 
as post-AGB candidates. Obviously, further infrared observations are
required to ascertain the true nature of these objects without optical
counterpart.

There are 106 evolved sources in the SGM catalogue located south of
declination $-$32$\deg$, which are the focus of this paper.

By the time the observations described in Sec.\ \ref{observations} 
were performed, we became aware of the
paper by \citet{deacon07}, who carried out a survey for water masers
in post-AGB candidates with OH 1662 MHz maser emission. Their survey
included 7 of the sources selected for our observations: IRAS
14341$-$6211, IRAS 17088$-$4221, IRAS 17164$-$3226, IRAS 17168$-$3736,
IRAS 17245$-$3951, IRAS 17310$-$3432, IRAS 17370$-$3357, with
one water maser
detection among them (IRAS 17088$-$4221). Therefore, these 7 sources
were explicitly left out of our observations, although we will include
them in the discussion of this paper (see
Sec. \ref{discussion}), whenever appropriate, because they fulfil the criteria used to select
the sample.

\section{Observations}
\label{observations}

We observed the $6_{16}\rightarrow 5_{23}$ transition of the water
molecule (rest frequency 22235.080 MHz), using the 64m antenna of the
Parkes Observatory\footnote{The Parkes telescope is part of the
  Australia Telescope, which is funded by the Commonwealth of
  Australia for operation as a National Facility managed by CSIRO}, on
2007 February 08-12. At this frequency, the half-power beamwidth of
the telescope is $\simeq  1\farcm 3$. The 1.3 cm receiver of this antenna
comprises a cryogenic High Mobility Electron Transistor. Both right
and left circular polarisation were observed simultaneously. As a
backend, we used the Multibeam Correlator, covering a bandwidth of 64
MHz (i.e., velocity coverage $\simeq 862.9$ km~s$^{-1}$) with 2048
spectral channels for each polarisation, thus yielding a spectral
resolution of 31.25 kHz (0.42 km~s$^{-1}$). Since little information
about the radial velocity of these sources is available, all observed
spectra were centred at $v_{\rm LSR} = 0$ km s$^{-1}$. The
observations were taken in position-switching mode with a total
integration time (on+off) of 30 min. The rms pointing accuracy of the
telescope was $\simeq 10''$.

Taking into account the detections found in Paper I, the
priority of the observations was set (in decreasing order) as: 
optically obscured sources, PNe, transition sources, and finally the
post-AGB stars. A total of 74 objects out of the 106 target sources
were observed.  Table \ref{observadas} gives the list of observed
  objects. The target coordinates were taken from  SGM, who list
  2MASS positions when a 
near-IR counterpart of the IRAS sources was identified. When SGM does
not provide these improved coordinates, we used those in the IRAS
Point Source Catalog. The observed sources are
divided in the following categories, according to the classification
in SGM:

\begin{itemize}
\item 23 PNe with optical counterpart
\item 8 transition objects with optical counterpart
\item 15 post-AGB stars with optical counterpart
\item 28 sources without optical counterpart (6 PNe candidates, 22 post-AGB
candidates)
\end{itemize}

These 74 sources plus the 7 objects observed by \citet{deacon07} (1 PN
and 4
post-AGB stars with optical counterpart, and 2 post-AGB candidates without
optical counterpart) comprised all objects without optical
counterpart as well as all PNe and transition objects with optical
counterpart included in the selected sample of 106 sources from SGM.

\begin{figure*}
\includegraphics[scale=1]{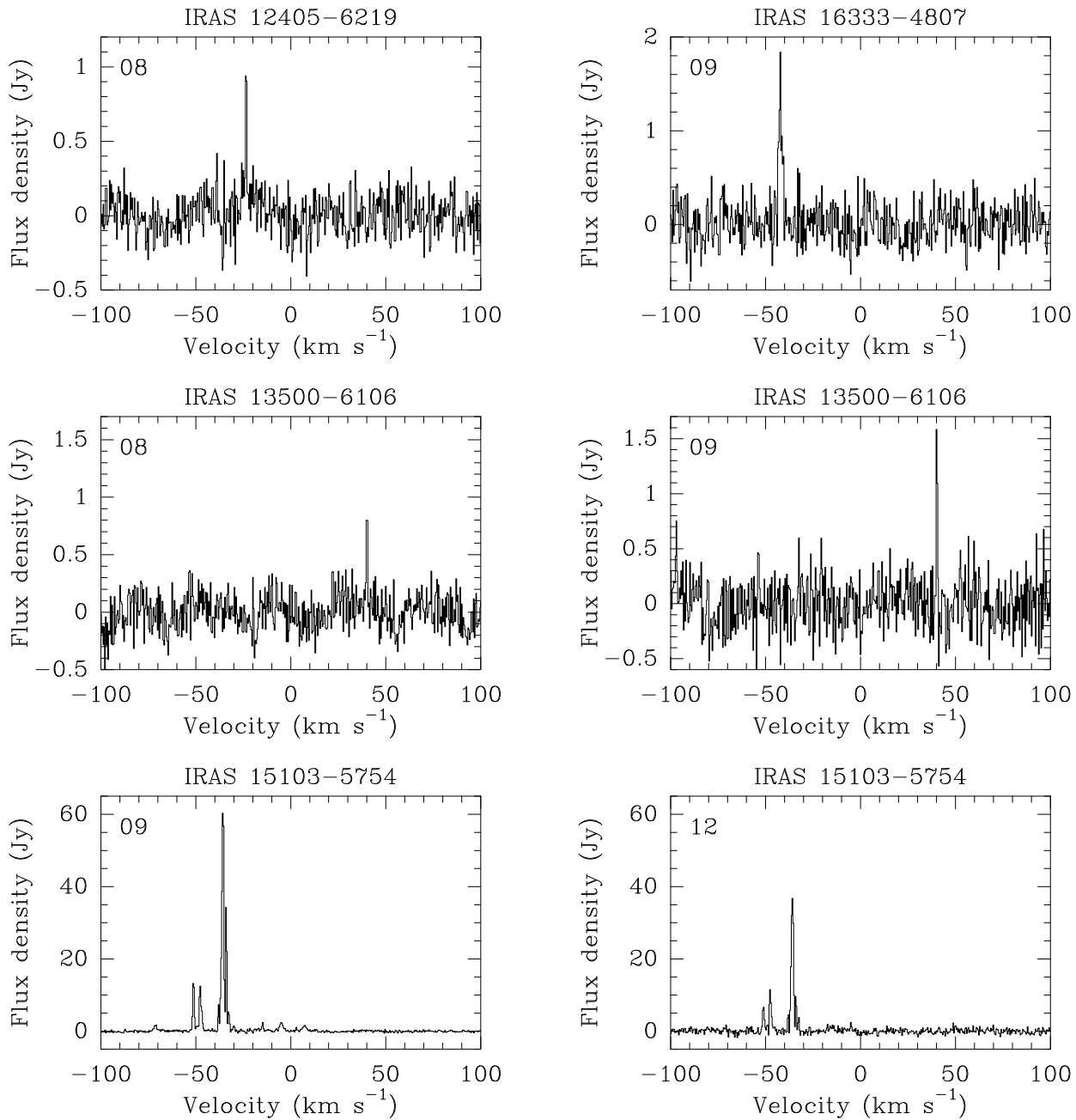}
\caption{Water maser spectra of the detected sources. The numbers shown at the upper left of the graph correspond to the day of observation, in 2007 February.}
\label{espectros}
\end{figure*}

\section{Results}
\label{results}

Our results are summarised in Table \ref{observadas}, where we list
all the observed sources, and Table \ref{detections}, where we give
the parameters of the
detections obtained. Four water masers were found during these
observations, all of them reported here for the first time: IRAS
12405$-$6219, IRAS 13500$-$6106, IRAS 15103$-$5754, and IRAS
16333$-$4807.  Their spectra are shown in Figs.~\ref{espectros} and
\ref{15103close}. All of them correspond to objects classified as
``sources without optical counterpart'' by SGM.

The sources IRAS 12405$-$6219, IRAS 15103$-$5754, and IRAS
16333$-$4807 have been classified by \citet{vdsteene93} as PN
candidates due to the presence of radio continuum emission, while IRAS
13500$-$6106 has been classified as a post-AGB star by \citet{gl97},
based on its position in the IRAS two-colour diagram and on its low
IRAS variability index (v=29). 

The association of the discovered water masers towards IRAS
12405$-$6219, IRAS 15103$-$5754 and IRAS 16333$-$4807 with the
radio continuum detected by \citet{vdsteene93} should be carefully
studied. The beam of our water maser observations is
$1\farcm 3$, while that of the radio continuum observations
is $4\farcs 5$. An unambiguous identification of the water
maser emission with the radio continuum sources requires higher
angular 
resolution observations of the water masers to derive a more accurate
position (see, e.g., \citealt{gomez08}).

In the following we will describe in detail the four detected objects
and comment on their nature. In order to identify possible candidates
to the maser-exciting source, we have also inspected MSX and 2MASS data.

\subsection{Detected sources}

\subsubsection{IRAS 12405$-$6219}

We detected water maser emission with a single spectral component
at $V_{\rm LSR} \simeq -23.6$ km s$^{-1}$ and a flux density of
$\sim$1~Jy (Fig.~\ref{espectros} and  Table~\ref{detections}).
A source of continuum emission at 6 cm was detected by
\citet{vdsteene93} at R.A.(J2000)=$12^h43^m32.04^s$,
Dec.(J2000)=$-62\deg36'13.9''$, with a flux density $S(6cm) = 14$ mJy.
This radio source is located $6''$ from the IRAS nominal position.

There is only one MSX source in the
beam of the water maser observations. This source (MSX6C
G302.0213+00.2542) is located $\sim6''$
from the reported radio continuum position, and is coincident with the
2MASS source J12433151-6236135. We can consider that the radio continuum
is associated with the IR source within the errors. The water maser
position is also probably associated with this source.


Although \citet{vdsteene93} classify it as a possible PN, this source
also fulfils the criteria used by \citet{hughes89} to select HII regions
from the IRAS Point Source Catalogue (log~S$_{60\mu\rm
  m}$/log~S$_{25\mu\rm m}~\ge$~0.25 and log~S$_{25\mu\rm
  m}$/log~S$_{12\mu\rm m}~\ge$~0.4). We will thus maintain
its classification as PN as doubtful.

\subsubsection{IRAS 13500$-$6106}

Our water maser spectrum shows one component with a maximum flux
density of $\sim$~1.5~Jy at $V_{\rm LSR} \simeq 40.0$~km~s$^{-1}$ (see
Fig.~\ref{espectros} and Table~\ref{detections}). This source was
classified as a post-AGB star by \citet{gl97} based on its IR colours.
The only MSX source in the beam of the water maser observations is
MSX6C G310.3099+00.6306, and there is no 2MASS source associated
with it.  No OH has been detected in this source by
\citet{telintel91}.

\subsubsection{IRAS 15103$-$5754}

\begin{figure}[t!]
\includegraphics[width=6cm,angle=-90]{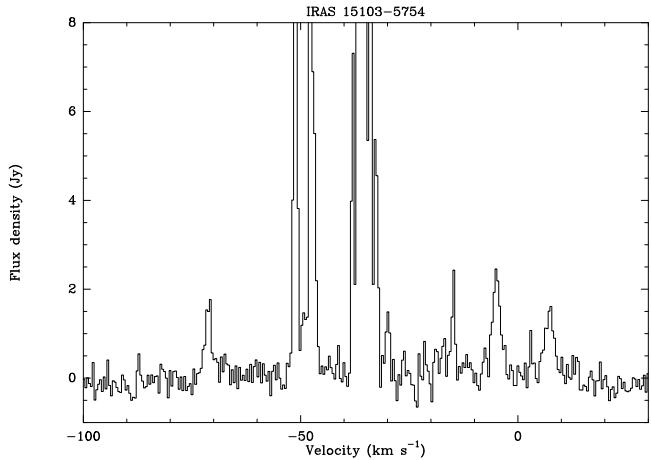}
\caption{Close up view of the water maser spectrum taken on 2007
  February 9 toward the PN candidate
  IRAS 15103-5754.}
\label{15103close}
\end{figure}

We have detected water maser emission with a maximum flux density of
$\sim$60~Jy (Fig. \ref{espectros} and
Table~\ref{detections}). The spectrum shows a rich set of multiple
components spanning $\simeq 80$ km~s$^{-1}$, from $V_{\rm LSR} \simeq
-70$ to $+10$ km~s$^{-1}$. In Fig.  \ref{15103close} we show a close
up view of the spectrum where we can see the weakest components. 
The intensity of the components shows significant variability (up to a
factor of $\simeq 3$) in the
spectra taken on two different days.

The radio continuum source reported by \citet{vdsteene93} is located
at R.A.(J2000)=$15^h14^m18.53^s$, Dec.(J2000)=$-58\deg05'21.1''$,
i.e., $15''$ from the IRAS nominal position, with a flux
density $S(6cm) = 115$ mJy. OH maser emission at 1612 MHz 
has been detected in this source
\citep{telintel96} at $V_{\rm LSR} \simeq -47$ km~s$^{-1}$ with a flux
density $\sim$0.3~Jy.
The only MSX source in the beam of our observations is MSX6C
G320.9062-00.2928, that is coincident with the 2MASS source
J15141845-5805203 (within $\sim 1\farcs 2$). The MSX
source is also coincident with the radio continuum position (within
$\sim 1\farcs 4$). These results suggest that IRAS
15103$-$5754 as a very likely PN, associated with the mentioned IR
sources.

The high velocity range found in the water maser components makes this
source a new candidate to the water fountain class. Moreover, if its
classification as a PN and its association with the water maser
emission are confirmed, this source would be the first PN showing water
fountain characteristics. The existence of ``water fountain-PNe'' was
suggested by \citet{gomez08}. They discussed the possibility of
finding a PN that would still maintain the high-velocity water masers
present during its previous post-AGB$-$water fountain stage. These
objects should be among the youngest and more massive PNe known, and
they would be excellent targets to study the ``real-time'' evolution
of a PN, since the time scales of evolution of this type of sources
would be very short and can be studied by means of high angular resolution
observations of the water masers.

\subsubsection{IRAS 16333$-$4807}

We have detected water maser emission towards this source (see
Fig.~\ref{espectros} and Table~\ref{detections}) with a maximum flux
density of $\sim$1.9~Jy at $V_{\rm LSR} \simeq 45$ km~s$^{-1}$.  The
radio continuum source detected by \citet{vdsteene93} is located at
R.A.(J2000)=$16^h37^m06.79^s$, Dec.(J2000)=$-48\deg13'42.6''$,
displaced $\simeq 7''$
from the IRAS nominal position, with a flux density $S(6cm) = 32.5$
mJy.

The only MSX source in the beam of our observations is MSX6C
G336.6445-00.6953, that is coincident with the source found by
\citet{vdsteene93} (within $\sim 1\farcs 4$) and with
the 2MASS source J16370659-4813429 (within
$\sim 1\farcs 5$). The likely association of the radio
continuum and IR sources makes this object a plausible PN candidate.

\subsection{Results from the two surveys for water masers in SGM sources}

The 179 sources observed in both campaigns (Robledo [Paper I] and
Parkes [this paper]) are distributed in evolutionary classes, according
to SGM, as follows:

\begin{itemize}

  \item 76 post-AGB stars with optical counterpart (42\%)

  \item 21 transition sources (12\%)

  \item 35 PN with optical counterpart (20\%)

  \item 47 sources without optical counterpart (2 HII regions, 35
    post-AGB candidates, 10 PN candidates) (26\%)

\end{itemize}

And the detections have been found towards:

\begin{itemize}

  \item 1 post-AGB star with optical counterpart: IRAS 07331+0021

  \item 3 post-AGB candidates without optical counterpart, one of them a
    ``water fountain'': IRAS 16552$-$3050 (water fountain), IRAS
    13500$-$6106, and IRAS 17580$-$3111.

  \item 1 confirmed PN with optical counterpart: IRAS 18061$-$2505.

  \item 3 PNe candidates without optical counterpart, one of them a possible
    ``water fountain PN'': IRAS 12405$-$6219, IRAS 15103$-$5754 (water
    fountain PN candidate), and IRAS 16333$-$4807.

  \item 1 AGB star without optical counterpart (OH/IR): IRAS 17443$-$2949

\end{itemize}

We note that IRAS 17580$-$3111 and IRAS 17443$-$2949 were initially
classified as obscured PNe with OH emission (OHPN) by
\citet{zijlstra89}, and as such were cited in \citet{suarez07}, but
\citet{gomez08} showed that the radio continuum reported by
\citet{ratag90} was not associated with these sources, and
\citet{garciahernandez07} classified them as post-AGB and AGB stars
respectively. We will adopt the classification of
\citet{garciahernandez07} for the discussion.

In summary, out of the nine detected sources, only one turned out to
be a ``classical'' post-AGB star with an optical counterpart. The rest
of the possible post-AGB stars in wich H$_2$O emission was found
do not have a
counterpart in the visible domain. The fact that these stars
could already be in
the post-AGB stage while still having a thick envelope that
prevents their 
detection in the optical, suggests that they are relatively massive
stars that have quickly evolved from the AGB. The new detections found
in this campaign confirm the trend already found in Paper I, relating
the presence of water masers in post-AGB stars or PN to
optically obscured sources.

Apparently, the incidence of water maser emission during the post-AGB
  phase is higher in obscured objects, i.e. those with more massive
  envelopes. Of  the 80 visible post-AGB stars observed, only 1 turned
  out to have 
water maser emission ($\sim$1\%), while we have found this emission in 4 out of 35 post-AGB star candidates, without 
optical counterpart ($\sim$11\%). This  
suggests that the presence of massive envelopes favour the pumping of
water masers during the post-AGB stage. Morever, our water maser
detections towards 
three obscured PN candidates might suggest a similar trend during the
PN phase. The detection of dense molecular gas 
toward an H$_2$O-PN \citep{tafoya07} also supports the existence of
massive envelopes in these objects.

The two most interesting results of our two surveys are, on one hand, the
detection of several new possible PNe with water maser emission (three
candidates found in this paper), and on the other hand, the
possibility of having found the first ``water fountain-PN''. In fact,
if the association of the water masers with the reported radio
continuum emission is confirmed in the three candidates found in this
paper (IRAS 12405$-$6219, IRAS 15103$-$5754, and IRAS 16333$-$4807), 
the number of PNe known to harbour water maser emission would
increase from 3 to 6. We expect these new H$_2$O-PNe to be bipolar, as
is the case for the other three known to belong to this group. High
resolution studies of these objects would be necessary to confirm this
hypothesis.
Furthermore, the confirmation of IRAS 15103$-$5754 as the first
``water fountain-PN'' would lead us into a new class of objects
susceptible of being one of the youngest and more massive PNe known.

\section{Discussion}
\label{discussion}

The original purpose of this work was to study the connection between
the presence of water masers and the evolution of the stars in the
post-AGB stage. This was the reason why the objects to be observed
were selected from the  SGM catalogue, since this catalogue contains the
largest number of post-AGB stars with spectroscopic classification
compiled to date. However,
as mentioned in Paper I, we 
have to be careful in deriving conclusions from the 
H$_2$O maser detection rate obtained in objects without optical counterpart,
since the SGM catalogue is incomplete for optically obscured sources.
However, the fact that only 26\% of the total sample was composed of
obscured objects, but 78\% of the detections have been found in such
sources, strongly suggests that the presence of massive envelopes 
favour the pumping of water masers in the post-AGB or PN stages.

\subsection{Position of the detected sources in the IRAS and MSX colour-colour diagram}

In order to study the evolutionary status of the sources associated with
water masers, we plotted them in the IRAS [12]-[25] {\it vs}
[25]-[60] diagram and in the MSX [8]-[12] {\it vs} [15]-[21] diagram.

\begin{figure}
\includegraphics[scale=0.4]{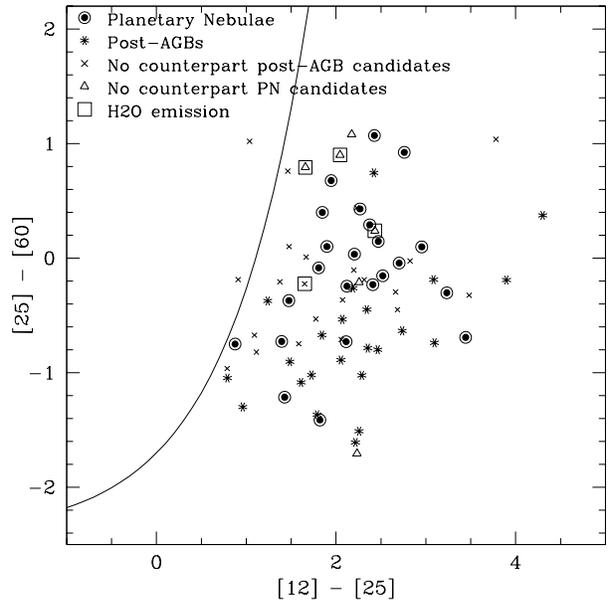}
\caption{ IRAS two colour diagram with the position of the observed
  and detected sources in the Parkes survey. The solid line is the
  curve modelled by \citet{bedijn} showing the position of the AGB
  stars.}
\label{completohs}
\end{figure}

\begin{figure}
\includegraphics[scale=0.4]{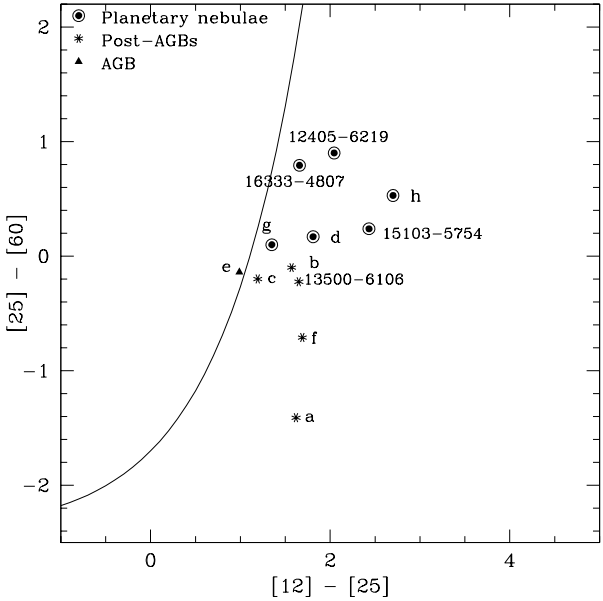}
\caption{ IRAS two colour diagram with the position of the detected
  sources in both campaigns (Robledo and Parkes detections). We have
  also included the position of the other two H$_2$O-PN known (IRAS
  17347$-$3139, and K3$-$35), and of IRAS 17088$-$4221. The sources detected in the present
  campaign are shown with the IRAS names, while the rest of the
  sources are identified by labels that correspond to: 
  a. IRAS 07331+0021, b. IRAS 16552$-$3050, c. IRAS 17088$-$4221, d.
  IRAS 17347$-$3139, e. IRAS 17443$-$2949, f. IRAS 17580$-$3111,
  g. IRAS 18061$-$2505
  h. K3$-$35.
  The
  solid line is the curve modelled by \citet{bedijn} showing the
  position of the AGB stars. Note that in this figure, we are using
  the same symbols for confirmed and candidate sources in each stage.}
\label{detecciones}
\end{figure}

Fig.~\ref{completohs} shows the position in the IRAS two-colour
diagram of all the 74 sources observed in this paper. We use different
symbols for the different types of sources and we also mark those
objects where water masers have been found.  In Fig.~\ref{detecciones}
we show the position in the IRAS two-colour diagram of the 9 sources
with water masers detected in both surveys (Robledo and Parkes). We
also include the post-AGB candidate IRAS 17088$-$4221, detected by
\citet{deacon07} and, for completeness, we also include the position
of the other two H$_2$O-PN: IRAS 17347$-$3139 \citep{degregorio04} and
K3$-$35 \citep{miranda01}. In this figure, we have used the same
symbol for confirmed and candidate water-maser-emitting PNe, and
analogously for confirmed and candidate post-AGB stars.

There seems to be a trend in Fig.~\ref{detecciones}, with confirmed
and candidate H$_2$O-PNe
having a larger value of [25]-[60] colour than water-maser-emitting post-AGB stars and
candidates. However, a confirmation of this trend will require a more
complete sample, specially for optically obscured evolved stars and,
of course, a conclusive determination of the association of masers
with the candidate objects, and of the true nature of these. 

\begin{figure}[t!]
\includegraphics[scale=0.45]{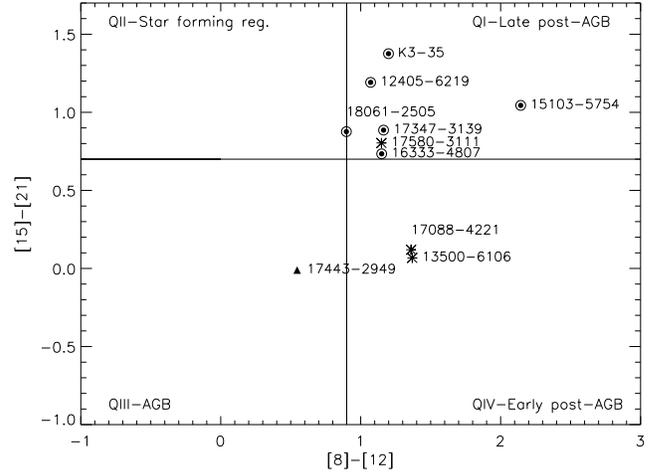}
\caption{MSX two colour diagram with the position of the detected
  sources in both campaigns (Robledo and Parkes detections). We have
  also included the position of the other two H$_2$O-PN known (IRAS
  17347$-$3139 and K3$-$35), and of IRAS 17088$-$4221. The symbols are the same as those defined
  in Fig.~\ref{detecciones}.}
\label{msx}
\end{figure}

In Fig.~\ref{msx} we show the position in the [8]$-$[12] {\it vs}
[15]$-$[21] MSX two-colour diagram of all the detections in both
surveys that have been observed with the MSX satellite (all but IRAS
07331+0021 and IRAS 16552$-$3050), plus of the other two H$_2$O-PNe
and the post-AGB candidate IRAS 17088$-$4221.
This diagram has been divided in four quadrants (QI, QII, QIII, QIV)
by \citet{sevenster02a} in her study of the position of
OH-maser-emitting AGB and post-AGB sources. The four quadrants
correspond to the position of classical AGB stars (QIII, low
left), early post-AGB stars (QIV, low right), more evolved
objects (QI, up right), and star forming regions (QII, up left).

Although the defined quadrants give only a tentative classification of
the sources located in each of them (e.g., we can also find post-AGB
stars and PNe in QII and QIII, see \citealt{suareztesis}), we note that
IRAS 17443$-$2949 is the only source that occupies its place in QIII,
the AGB quadrant, which supports its identification by
\citet{garciahernandez07} as an AGB star. All the PNe and PNe
candidates are located in the region QI, occupied by evolved post-AGB
and PNe, and the only objects located in the region QIV (early post-AGB
stars) are the post-AGB candidates IRAS 13500$-$6106 and IRAS 17088$-$4221. Since there are no
objects in QII, it is likely that none of our detections is a
star-forming region, even if we maintain our doubts in the PN nature
of IRAS 12405$-$6219.

\subsection{Evolutionary consequences of the presence of water masers
  in post-AGB stars and PNe}

As was already stated in \citet{lewis89} and \citet{gomezy90}, the
conditions of H$_2$O maser pumping in the AGB phase can continue
during the post-AGB stage for at most $\sim$100 yr in the case of a spherical
mass-loss. Therefore, this time scale indicates that most
  H$_2$O masers found in  post-AGB stars are probably originated during this phase.


As it was suggested in Paper I and in
\citet{gomez08}, the presence of asymmetrical mass-loss in the
post-AGB stage can favour the necessary conditions to pump water
masers in this phase and in that of PN. 
This is independent of the previous presence or not of these masers
during the AGB phase, as it is the case of the water fountains, such
as IRAS 16552$-$3050 \citep{suarez08}, first detected in Paper I. If the three
new obscured PNe candidates found to harbour water masers in this
paper turn out to have bipolar morphologies, as the other three
H$_2$O-PNe known, it will help to support the suggestion made in
Paper I and in \citet{gomez08} relating the presence of water masers
in the PN stage (the H$_2$O-PNe) to the phenomenon of the water
fountains. Moreover, the discovery in this survey of the PN candidate
IRAS 15103$-$5754 that shows ``water fountain'' characteristics would
be an excellent evidence of the adequacy of the evolutionary model
proposed, since its existence was predicted in \citet{gomez08}. This
source, if definitively confirmed as a ``water fountain-PN'', would be
the missing link between the ``water fountains'' and the H$_2$O-PNe.

A proposed scenario for the maser emission in the ``water
fountains'' \citep{imai07} suggests that after the water fountain is
formed, the bipolar jets that harbour the water masers will reach the
region of the former AGB envelope where the density would not be high
enough to maintain the pumping of the masers. However, as discussed in
\citet{gomez08}, the regions close to the equator of the star, could
still fulfil the adequate conditions to pump the water masers, even
when the central star has reached enough temperature to become a PN.
If the star is massive enough to evolve rapidly through these stages,
we still could see the water fountain when it has reached the PN
stage. This could be the case of IRAS 15103$-$5754, if high resolution
observations confirm that the high velocity maser components arise from
bipolar jets.

K3$-$35 would be in the next stage, in which the jets
would have been slowed down before the quenching of the water masers
in the lobes of the PN. The rest of the H$_2$O-PNe detected up to now
would correspond to the final stage when the masers are only pumped in
dense areas close to the central stars.

Following the discussion in \citet{gomez08}, we suggest a tentative scenario
and characteristics for the evolution of the water masers in the
post-AGB and PN stages: 

\begin{itemize}

\item Post-AGB stars with water maser emission that does not
    trace high velocity jets (e.g. IRAS
    07331+0021). In a few cases, this maser emission might be the
    remnant of the one pumped during the AGB. In others, however, 
    the emission could have been originated during the post-AGB phase
    itself, but the source characteristics  do not favor
    maser amplification in jets, or do not even allow the
    formation of a jet.

\item Post-AGB stars that show water fountain characteristics
  (e.g. IRAS 16552$-$3050; 11 sources known, 9 of them obscured in the
  optical). 
\item Young bipolar PN with water fountain characteristics (e.g. IRAS
  15103$-$5754; this is the only candidate known up to now and it is
  obscured in the optical range). 

\item Young bipolar PN showing water masers at the tips of the bipolar
  lobes and at close to the central star (e.g. PN K3$-$35; this is the
  only source that has shown in one epoch these characteristics).

\item Young bipolar PN showing water masers only close to the central
  star (e.g. IRAS 18061$-$2505; 2 sources known, one of them without
  optical counterpart, and 2 candidates also obscured in the optical).

\end{itemize}


 While this scenario could be valid in general, we note that
  geometrical effects could introduce some caveats when trying to fit
  a particular object into it. For instance, the orientation of the
  object with respect to the observer affects the observational
  appearance of its maser emission. A source with a jet of the ``water
  fountain'' type would never show up the typical high-velocities if
  the jet is close to the plane of the sky.

In view of the limited number of objects in each of the proposed
stages, it is evident that greater efforts should be done to find
obscured evolved objects harbouring water masers. The possibility of
finding among them new "water fountains" (post-AGB or PN) or new
H$_2$O-PNe seems to be higher for obscured objects and these studies would
contribute to confirm 
or reject the proposed evolutionary scenario, and to constraint
the time spent 
by the stars in each stage.

\section{Conclusions}
\label{conclusions}

We have performed a survey for water masers in a sample of 74 evolved
stars selected from the SGM atlas, using the Parkes 64-m antenna.

We have found four new detections of water masers, three of them in
obscured PNe candidates: IRAS 12405$-$6219, IRAS 15103$-$5754, and
IRAS 16333$-$4807; and one in an obscured post-AGB candidate: IRAS
13500$-$6106. This could increase the number of H$_2$O-PN known from
three to six.

One of the PN candidates where water masers have been found, IRAS
15103$-$5754, has also characteristics of a water fountain. If its PN
nature and its association with water maser emission is confirmed, it
would be the first known case of a "water fountain" PN.  

We propose an evolutionary scenario for water masers in the post-AGB
and PNe stages, in which we try to accommodate all the detections in
our survey. After a period in which a post-AGB star could maintain the
remaining of the water masers pumped during the AGB, some of them
develop water fountain characteristics. The ``water fountain'' masers
could still be present
in very young bipolar PNe. Later in the PN stage, lower-velocity water
masers would be first present both along jets and close to the
central object, with only the central masers remaining in more
evolved PNe.

In this proposed 
scenario for the evolution of the water masers in the post-AGB and PN
stages,  IRAS
15103$-$5754 could represent the "missing link" between
water fountains and H$_2$O-PNe.

We confirm the tendency suggested in Paper I that the presence of
water masers is favoured in obscured sources with massive envelopes.
\begin{acknowledgements}
  We would like to thank our referee, Dr. Dieter Engels, for his
  careful and useful review.  O.S., J.F.G., J.M.T., and G.A. are
  partially supported by Ministerio de Ciencia e Innovaci\'on, grants
  AYA2008-06189-C03 and AYA2005-08523-C03 (co-funded with FEDER
  funds).  O.S. and L.F.M. are partially supported by Ministerio de
  Ciencia e Innovaci\'on, grant AYA2008-01934. O.S., J.F.G., L.F.M.,
  J.M.T., and G.A.  also acknowledge support from Consejer\'{\i}a de
  Innovaci\'on, Ciencia y Empresa of Junta de Andaluc\'{\i}a. This
  publication makes use of data products from the Two Micron All Sky
  Survey, which is a joint project of the University of Massachusetts
  and the Infrared Processing and Analysis Center/California Institute
  of Technology, funded by the National Aeronautics and Space
  Administration and the National Science Foundation.
\end{acknowledgements}



\longtab{1}{
\begin{longtable}{cccclcc}
\caption{Table of observed sources. The first day of observation 2007-FEB-08 correspond to the Julian Date 2454139.5.  \label{observadas}}\\
\hline \hline 
IRAS name & R.A. (J2000) &  Dec. (J2000)& Observation date & rms (Jy) & Classification &  Notes\\
& & & & & \citet{suarez06} & \\
\hline
\endfirsthead
\caption[]{Observed sources (continued)}\\
\hline\hline
IRAS name & R.A. (J2000) &  Dec. (J2000)& Observation date & rms (Jy)
& Classification & Notes \\
& & & & & \citet{suarez06} & \\
\hline
\endhead
\hline
\endlastfoot

07027-7934 &    06:59:26.4 &    -79:38:47 & 2007-FEB-09 & 0.22 & PN \\  
08046-3844 &    08:06:28.4 &    -38:53:24 & 2007-FEB-10 & 0.19 & PN\\
08242-3828 &    08:26:03.8 &    -38:38:48 & 2007-FEB-08 & 0.19 & No optical counterpart & 1\\
08281-4850 &    08:29:40.6 &    -49:00:04 & 2007-FEB-12 & 0.21 & Post-AGB \\
08351-4634 &    08:36:45.8 &    -46:44:46 & 2007-FEB-08 & 0.22 & No optical counterpart & \\
08355-4027 &    08:37:24.7 &    -40:38:04 & 2007-FEB-10 & 0.20 & PN \\  
08418-4843 &    08:43:29.5 &    -48:54:47 & 2007-FEB-10 & 0.17 & PN \\  
08574-5011 &    08:59:02.3 &    -50:23:40 & 2007-FEB-10 & 0.17 & PN \\
09362-5413 &    09:37:51.8 &    -54:27:09 & 2007-FEB-09 & 0.60 & PN \\  
09370-4826 &    09:38:53.3 &    -48:40:10 & 2007-FEB-08 & 0.25 & No optical counterpart &  \\
09425-6040 &    09:44:01.7 &    -60:54:26 & 2007-FEB-12 & 0.23 & Post-AGB \\
09500-5236 &    09:51:49.2 &    -52:50:53 & 2007-FEB-08 & 0.22 & No optical counterpart &  \\
09517-5438 &    09:53:27.1 &    -54:52:40 & 2007-FEB-09 & 0.20 & PN \\  
10029-5553 &    10:04:40.1 &    -56:08:37 & 2007-FEB-09 & 0.21 & PN \\
10115-5640 &    10:13:19.7 &    -56:55:32 & 2007-FEB-09 & 0.22 & PN \\
10178-5958 &    10:19:32.5 &    -60:13:29 & 2007-FEB-10 & 0.17 & Transition \\
10197-5750 &    10:21:33.9 &    -58:05:48 & 2007-FEB-11 & 0.23 & Transition \\
10215-5916 &    10:23:19.5 &    -59:32:05 & 2007-FEB-11 & 0.25 & Transition \\
10256-5628 &    10:27:35.2 &    -56:44:20 & 2007-FEB-11 & 0.19 & Post-AGB \\
11201-6545 &    11:22:18.9 &    -66:01:51 & 2007-FEB-11 & 0.20 & Post-AGB \\
11339-6004 &    11:36:20.7 &    -60:20:53 & 2007-FEB-08 & 0.18 & No optical counterpart &  \\
11353-6037 &    11:37:42.9 &    -60:53:51 & 2007-FEB-11 & 0.22 & Transition \\ 
11381-6401 &    11:40:32.0 &    -64:18:35 & 2007-FEB-08 & 0.17 & No optical counterpart & \\
11387-6113 &    11:41:08.7 &    -61:30:17 & 2007-FEB-11 & 0.22 & Post-AGB \\ 
11531-6111 &    11:55:38.0 &    -61:28:17 & 2007-FEB-11 & 0.22 & Transition \\ 
12145-5834 &    12:17:16.1 &    -58:51:30 & 2007-FEB-12 & 0.24 & Post-AGB \\
12262-6417 &    12:29:04.2 &    -64:33:37 & 2007-FEB-08 & 0.18 & No optical counterpart & \\
12302-6317 &    12:33:07.0 &    -63:33:43 & 2007-FEB-12 & 0.23 & Post-AGB \\
12309-5928 &    12:33:44.6 &    -59:45:19 & 2007-FEB-08 & 0.18 & No optical counterpart \\
12316-6401 &    12:34:36.0 &    -64:18:17 & 2007-FEB-09 & 0.23 & PN \\
12405-6219 &    12:43:31.5 &    -62:36:14 & 2007-FEB-08 & 0.12 & No optical counterpart & 2\\
13203-5917 &    13:23:32.2 &    -59:32:50 & 2007-FEB-12 & 0.24 & Post-AGB \\
13293-6000 &    13:32:39.2 &    -60:15:39 & 2007-FEB-08 & 0.18 & No optical counterpart \\
13356-6249 &    13:39:05.8 &    -63:04:44 & 2007-FEB-08 & 0.20 & No optical counterpart & \\
13416-6243 &    13:45:07.3 &    -62:58:17 & 2007-FEB-12 & 0.24 & Post-AGB \\
13427-6531 &    13:46:25.7 &    -65:46:24 & 2007-FEB-08 & 0.20 & No optical counterpart \\
13500-6106 &    13:53:34.5 &    -61:20:52 & 2007-FEB-08 & 0.15 & No optical counterpart & \\
           &               &              & 2007-FEB-09 & 0.23 \\
13529-5934 &    13:56:24.6 &    -59:48:57 & 2007-FEB-08 & 0.22 & No optical counterpart & \\
14079-6402 &    14:11:46.3 &    -64:16:24 & 2007-FEB-09 & 0.22 & PN \\
14122-5947 &    14:15:53.3 &    -60:01:38 & 2007-FEB-10 & 0.13 & PN \\
           &               &              & 2007-FEB-11 & 0.22 \\
14177-5824 &    14:21:19.9 &    -58:38:22 & 2007-FEB-10 & 0.19 & PN \\ 
14345-5858 &    14:38:20.0 &    -59:11:46 & 2007-FEB-09 & 0.22 & PN \\
14482-5725 &    14:51:57.3 &    -57:38:19 & 2007-FEB-11 & 0.19 & Post-AGB \\
14488-5405 &    14:52:28.7 &    -54:17:43 & 2007-FEB-11 & 0.20 & Post-AGB \\
15039-4806 &    15:07:27.4 &    -48:17:54 & 2007-FEB-11 & 0.20 & Post-AGB\\
15066-5532 &    15:10:26.0 &    -55:44:13 & 2007-FEB-11 & 0.20 & Transition \\
15093-5732 &    15:13:12.4 &    -57:43:40 & 2007-FEB-08 & 0.21 & \\
15103-5754 &    15:14:18.5 &    -58:05:20 & 2007-FEB-09 & 0.24 & No optical counterpart & 2\\
           &               &              & 2007-FEB-12 & 0.6 \\
15144-5812 &    15:18:21.9 &    -58:23:12 & 2007-FEB-09 & 0.22 & No optical counterpart & \\
15154-5258 &    15:19:08.2 &    -53:09:47 & 2007-FEB-10 & 0.17 & PN \\
15210-6554 &    15:25:31.7 &    -66:05:20 & 2007-FEB-11 & 0.19 & Post-AGB \\
15534-5422 &    15:57:21.1 &    -54:30:46 & 2007-FEB-09 & 0.21 & No optical counterpart \\
15559-5546 &    15:59:57.4 &    -55:55:34 & 2007-FEB-10 & 0.18 & PN\\
15579-5445 &    16:01:50.8 &    -54:53:40 & 2007-FEB-10 & 0.17 & PN \\
16053-5528 &    16:09:20.2 &    -55:36:10 & 2007-FEB-10 & 0.17 & PN \\
16114-4504 &    16:15:03.0 &    -45:11:54 & 2007-FEB-10 & 0.18 & PN \\
16228-5014 &    16:26:31.3 &    -50:21:27 & 2007-FEB-09 & 0.22 & No optical counterpart \\
16333-4807 &    16:37:06.1 &    -48:13:42 & 2007-FEB-09 & 0.20 & No optical counterpart & 2\\
16494-3930 &    16:52:55.4 &    -39:34:56 & 2007-FEB-10 & 0.20 & Post-AGB \\
16518-3425 &    16:55:08.4 &    -34:30:10 & 2007-FEB-09 & 0.23 & No optical counterpart &  \\
16529-4341 &    16:56:34.0 &    -43:46:15 & 2007-FEB-10 & 0.18 & PN \\
16584-3710 &    17:01:52.1 &    -37:14:54 & 2007-FEB-09 & 0.21 & No optical counterpart \\
16594-4656 &    17:03:11.2 &    -47:00:21 & 2007-FEB-10 & 0.19 & Transition \\
17009-4154 &    17:04:29.6 &    -41:58:39 & 2007-FEB-09 & 0.22 & No optical counterpart & \\
17010-3810 &    17:04:27.3 &    -38:14:42 & 2007-FEB-09 & 0.22 & No optical counterpart \\
17088-4227 &    17:12:21.8 &    -42:30:50 & 2007-FEB-10 & 0.18 & PN \\
17119-5926 &    17:16:21.1 &    -59:29:23 & 2007-FEB-10 & 0.17 & PN \\
17130-4029 &    17:16:29.0 &    -40:32:31 & 2007-FEB-09 & 0.20 & No optical counterpart & \\
17131-3330 &    17:16:26.2 &    -33:33:24 & 2007-FEB-10 & 0.21 & PN \\
17153-3814 &    17:18:44.7 &    -38:17:21 & 2007-FEB-09 & 0.21 & No optical counterpart & 3\\
17234-4008 &    17:26:56.1 &    -40:11:04 & 2007-FEB-09 & 0.22 & No optical counterpart & \\
17311-4924 &    17:35:02.5 &    -49:26:26 & 2007-FEB-10 & 0.18 & Transition \\
17418-3335 &    17:45:08.7 &    -33:36:06 & 2007-FEB-09 & 0.20 & No optical counterpart & 3\\
17476-4446 &    17:51:16.4 &    -44:47:29 & 2007-FEB-10 & 0.19 & Post-AGB \\         

\end{longtable}

\noindent
$^1$ OH maser emission detected by \citet{telintel91}. The spectrum
does not show the typical double-peaked profile of OH/IR stars, but
presents multiple components. This
suggest the source is a post-AGB star.\\
$^2$ Radio continuum emission detected by
  \citet{vdsteene93}. Thus, we consider this source as a ``PN candidate''.\\
$^3$  Radio continuum emission detected by \citet{ratag90}. Thus, we consider this source as a ``PN candidate''.\\

}

\begin{table*}
\begin{center}
\caption{Water maser detections \label{detections}}
\begin{tabular}{ccccccc}
\hline \hline 
IRAS  & $V_{\rm peak}$ & $S_{\rm peak}$ & $\int S_\nu d\nu$ & Date \\
          & (km s$^{-1}$)   & (Jy)                      & (Jy km s$^{-1}$)                 \\
\hline

12405-6219 & $-23.6\pm 0.4$ & $0.94\pm 0.24$ & $1.5\pm 0.4$ & 2007-FEB-08 \\
13500-6106 & $39.9 \pm 0.4$ & $0.8\pm 0.3$  & $1.2\pm 0.5$ & 2007-FEB-08 \\
                      & $40.0 \pm 0.4$ & $1.6\pm 0.5$  & $0.9\pm 0.7$ & 2007-FEB-09 \\
15103-5754 & $-35.8 \pm 0.4$ & $60.3 \pm 0.5$ & $179\pm 3$ & 2007-FEB-09 \\ 
                      & $-35.8 \pm 0.4$ & $36.7\pm 1.2$ & $95\pm 8$   & 2007-FEB-12 \\
16333-4807 & $-42.2 \pm 0.4$ & $1.8\pm 0.4$ & $3.5\pm 0.7$ &  2007-FEB-09 \\ 

\hline
\end{tabular}
\end{center}
\end{table*}

\end{document}